\documentclass[11pt,a4paper]{article}

\usepackage{jheppub}
\usepackage[T1]{fontenc}
\usepackage{amsmath,amssymb,amsfonts}
\usepackage{physics}
\usepackage{bm}
\usepackage{bbm}
\usepackage{slashed}
\usepackage{braket}
\usepackage{graphicx}
\usepackage{hyperref}
\usepackage{xcolor}

\title{\boldmath Effective Equations of Motion for Massive Higher Spins with Generic Gyromagnetic Ratio}

\author[a]{Karim Benakli}
\affiliation[a]{Sorbonne Universit\'e, CNRS, Laboratoire de Physique Th\'eorique et Hautes \'Energies, LPTHE, F-75005 Paris, France}
\emailAdd{kbenakli@lpthe.jussieu.fr}

\abstract{
We construct consistent effective equations of motion for massive charged particles of spin 2 and spin $3/2$ interacting with constant electromagnetic backgrounds. While tree-level unitarity restricts point-like particles to a universal gyromagnetic ratio $g=2$, long-lived composite states such as the $\Omega^-$ baryon generically deviate from this value. We propose describing such non-minimal couplings without triggering pathologies such as superluminal propagation or the loss of degrees of freedom by treating the system as an Effective Field Theory expanded in powers of the background field strength. By introducing specific operators, we show that algebraic consistency can be restored perturbatively for arbitrary $g$-factors.
}

\keywords{Higher--spin fields, Effective Field Theory, Electromagnetic backgrounds, Omega baryon}

\begin{document}

\maketitle
\flushbottom

\section{Introduction}
\label{sec:intro}
In 1964, a photographic plate from the Brookhaven 80-inch bubble chamber
recorded the first observation of the $\Omega^{-}$ hyperon~\cite{Barnes:1964pd}.
This baryon, with quark content $sss$ and strangeness $-3$, had been predicted
earlier by Gell-Mann and Ne'eman as the missing member of the
SU(3) flavour-symmetry decuplet within the ``eightfold way''. This discovery
provided crucial support for the quark-model classification.

The identification relied on the reconstruction of the weak-decay chain
\[
\Omega^- \to \Xi^0 \pi^- \;\to\; \Lambda\,\pi^0\,\pi^-
\;\to\; p\,\pi^-\,\pi^0\,\pi^-,
\]
with charged tracks visible in the bubble-chamber photograph.
The measured momenta and decay geometry implied a mass in agreement with the SU(3) prediction for the $\Omega^{-}$.
A curved trajectory of length $\sim 2.5$\,cm in the nearly uniform magnetic
field of the chamber was identified as the flight path of the $\Omega^{-}$.

Two features render this event particularly relevant for the present study.
First, the lifetime of the $\Omega^{-}$ ($\sim 10^{-10}\,$s) is sufficiently
long for it to behave as a quasi-stable relativistic particle over a
macroscopic distance in the bubble chamber, allowing its trajectory to be well
described by an effective equation of motion in a static magnetic field.
Second, although the spin was not determined experimentally in 1964, a
high-statistics angular-distribution analysis by the \textsc{BaBar}
collaboration decades later confirmed unambiguously that the $\Omega^{-}$
carries spin~$3/2$~\cite{BaBar:2006omx}.

Thus, the $\Omega^{-}$ constitutes one of the earliest direct observations of a
massive, charged relativistic baryon---today known to carry spin~$3/2$---propagating in
an approximately constant electromagnetic background. It provides a historically
concrete realisation of the theoretical question raised already by Dirac in
1936~\cite{Dirac:1936tg}. He wrote then:
«The elementary particles known to present-day physics, the electron, positron,
neutron, and proton, each have a spin of a half, and thus the work of the
present paper will have no immediate physical application. All the same, it is
desirable to have the equation ready for a possible future discovery of an
elementary particle with a spin greater than a half, or for approximate
application to composite particles. Further, the underlying theory is of
considerable mathematical interest.»

The relativistic description of massive higher-spin fields begins with
Fierz and Pauli~\cite{Fierz:1939ix}, who formulated the free equations of
motion and constraints for arbitrary spin $s$.
Higher-spin fields necessarily contain unphysical lower-spin components; a delicate
interplay of algebraic and differential constraints is required to eliminate
them and ensure propagation of the correct $2s+1$ physical degrees of freedom.
For free massive fields, these Fierz--Pauli systems have been shown to be
consistent and hyperbolic.

Simply introducing electromagnetic interactions through minimal coupling,
$\partial_m \rightarrow D_m=\partial_m+i e A_m$, proves to be quite problematic.
For spin~$\tfrac{3}{2}$, Johnson and Sudarshan showed that minimal coupling
leads to noncovariant equal-time commutators, signalling an inconsistency at
the canonical level~\cite{Johnson:1960vt}.
For both spin~$\tfrac{3}{2}$ and spin~2, Velo and Zwanziger demonstrated that
the minimally coupled Fierz--Pauli equations develop pathological
characteristic surfaces, leading either to superluminal propagation or to
loss of a constraint and propagation of an unphysical number of modes
\cite{Velo:1969bt,Velo:1969txo,Velo:1972rt}. The loss of a constraint appears at the
classical level at the same background field strength as the problem in
canonical quantization, indicating a common origin.
These problems were later traced to the degeneracy of the secondary
constraints in the presence of the external electromagnetic field.
It became clear that minimal coupling alone cannot yield consistent equations
of motion for massive charged higher-spin fields: appropriate nonminimal
interactions involving $F_{mn}$ are indispensable (see for example \cite{federbush_1961,Shamaly:1972zu,Hortacsu:1974bm,Madore:1975zz,Kobayashi:1978xd,Kobayashi:1978mv,Piccinini:1984dd,Buchbinder:2000fy,Deser:2001dt}).

For spin~2, the first massive level of the bosonic open string led Argyres and
Nappi to construct an explicit Lagrangian, giving a fully nonlinear system
consistent in constant electromagnetic backgrounds~\cite{Argyres:1989cu}. While
the original Lagrangian was only consistent in $D=26$ dimensions, it nevertheless leads
to a nonlinear Fierz--Pauli system that can be generalized to arbitrary lower
dimensions. Porrati, Rahman and Sagnotti subsequently showed how this consistent
system can be extended to all integer spins using the Virasoro constraints of
the free bosonic open string~\cite{Porrati:2010hm}.

It is remarkable that the equations of motion for the integer-spin fields are
quite simple. They could, in hindsight, have been guessed as a generalisation
of the well-known Proca equation for a spin-1 field, supplemented by the
requirement that the gyromagnetic ratio take the value $g=2$. This value is
required for good high-energy behaviour of a \emph{pointlike} higher-spin
particle: tree-level amplitudes remain unitary up to energies much larger than
the particle masses~\cite{Ferrara:1992yc}. The spin~$\tfrac{3}{2}$ case proved
far more subtle. Early attempts to guess a Fierz--Pauli system were not
successful. For instance, an example proposed in~\cite{Ferrara:1992yc}
was shown in~\cite{Porrati:2009bs} to require explicit lower-spin auxiliary
fields in a way that spoiled the simple higher-spin interpretation. A large
class of non-minimal couplings added to the free Rarita--Schwinger
Lagrangian failed to restore the full consistency conditions in an
electromagnetic background~\cite{Deser:2000dz}. An important step forward was
the proof by Porrati and Rahman that an \emph{infinite} tower of $F_{mn}$-dependent
terms, introduced recursively, can in principle yield a consistent
Fierz--Pauli system~\cite{Porrati:2009bs}. More recently, similar recursive
systems have been obtained without lower-spin fields in the fundamental
Lagrangian, but only with the help of auxiliary fields and without producing a
closed-form expression for the full nonlinear system~\cite{Delplanque:2024xst}.

For spin~$\tfrac{3}{2}$, a fully explicit and consistent system of equations
was constructed only recently. In a series of works, explicit equations of
motion and constraints for massive spin~2 and spin~$\tfrac{3}{2}$ were derived
directly in four dimensions
\cite{Benakli:2021jxs,Benakli:2022ofz,Benakli:2022edf,Benakli:2023aes,Benakli:2023qlg,Benakli:2024aes}.
The derivation originates from open superstring field theory in superspace,
where the lower-spin excitations appear as Lagrange multipliers.
Although intermediate expressions mix fields of different spin, the resulting
Fierz--Pauli systems can be rewritten in a basis that isolates the physical
spin degrees of freedom. For spin~2, this reproduces the Argyres--Nappi
structure directly in four dimensions
\cite{Benakli:2021jxs,Benakli:2022ofz,Benakli:2022edf,Benakli:2023aes}. The
spin–$\tfrac{3}{2}$ equations necessarily mix left- and right-handed components
in a nontrivial way that cannot be extrapolated from previously proposed
ansätze.

Both the integer- and half-integer–spin constructions that are known to be
consistent in constant electromagnetic backgrounds share a common feature:
they fix the gyromagnetic ratio to the universal value $g=2$, as expected for
an elementary pointlike particle. Composite states, however, are extended
objects. Their finite size implies an intrinsic ultraviolet (UV) cutoff set by
their mass, and the arguments enforcing $g=2$ in the point-particle limit no
longer apply. Electromagnetic couplings of such states naturally receive
non-universal contributions from higher multipoles, and one generically expects
$g\neq2$. A simple and physically relevant illustration is the $\Omega^{-}$
baryon. Its magnetic moment is measured to be
$\mu_{\Omega^-}=(-2.02\pm0.05)\,\mu_N$~\cite{ParticleDataGroup:2024cfk}. Using the standard spin–$3/2$ relation
\[
\mu = g\,\frac{Q}{2M}\,S ,
\]
with $Q=-e$, $S=3/2$, and $M_{\Omega^-}=1.672~\text{GeV}$, one finds
\[
g_{\Omega^-}\simeq 2.40\pm0.06 ,
\]
a clear deviation from the pointlike value $g=2$, consistent with its composite
nature.

This makes it clear that the existing constant-field generalizations of
Fierz--Pauli theory are not sufficient: a consistent framework accommodating
$g\neq2$ should exist, and is required for a realistic description of composite
higher-spin states such as the $\Omega^{-}$.

A controlled framework for describing deviations from the universal $g=2$
pointlike limit is provided by Effective Field Theory (EFT).  In this setting,
the higher-spin field is treated as a fundamental degree of freedom only below a
compositeness scale $\Lambda$, and its equations of motion are constructed as
\emph{systematic deformations} of the consistent $g=2$ systems.  These
deformations appear as a local operator expansion, suppressed both by inverse
powers of the heavy scale and by powers of the electromagnetic background field
strength.

For long-lived hadronic resonances such as the $\Omega^{-}$, the mass is set by
the same confining dynamics that determines the compositeness scale.  It is
therefore natural to identify $\Lambda\simeq M$ and to organize the EFT in
powers of $1/M$.  The expansion is governed by two small parameters,
\begin{equation}
\epsilon_p = \frac{|k|}{M}\ll1,
\qquad
\epsilon_F = \frac{|eF|}{M^{2}}\ll1,
\label{eq:EFT-small-intro}
\end{equation}
where $k^m=p^m-Mv^m$ is the residual momentum in the heavy-field decomposition
$\Phi = e^{-iMv\cdot x}\,\tilde\Phi$.  
Throughout this work the background field is taken to vary slowly along the
trajectory (constant, with $D_\ell F_{mn}=0$), ensuring that operators
containing derivatives of $F_{mn}$ are parametrically suppressed.

Two remarks clarify the interpretation of these parameters.

\smallskip
Although $F_{mn}$ is not a scalar, the EFT counting is fully Lorentz covariant.
“Powers of $F$’’ refer to the number of explicit insertions of $F_{mn}$ in local
operators, not to the magnitude of its components in a particular frame.  The
relevant small quantities are the invariants
\[
\epsilon_F \sim \frac{|e|\,\sqrt{F_{mn}F^{mn}}}{M^{2}}, 
\qquad
\epsilon_{\tilde F} \sim \frac{|e|\,\sqrt{F_{mn}\tilde F^{mn}}}{M^{2}},
\]
which in a purely magnetic field reduce to $|eB|/M^2$ and $0$.  Thus the $F$
expansion simply orders the Pauli-type (one $F$), quadrupole-type (two $F$),
etc.\ operators, each suppressed by additional powers of $\epsilon_F$.

\smallskip
Near the mass shell, the operator $\Delta=D^{2}+M^{2}$ acting on $\Phi$ gives
$\Delta\Phi\sim M k\,\Phi$, so residual derivatives scale as $k/M$.  In many
physical situations—for example, in Landau-level problems—one typically finds
$|k|\sim |eF|/M$, and therefore $\epsilon_p$ and $\epsilon_F$ are of comparable
size.  The EFT regime is characterized by
\[
\epsilon_F \ll 1,
\qquad
\epsilon_p \ll 1,
\]
which guarantees that derivative-suppressed operators and higher
electromagnetic multipoles are consistently organised, and that boosts (which
mix electric and magnetic components) do not spoil the smallness of the Lorentz
invariants entering the expansion.

\bigskip
For resonances arising from a confining interaction, the assumption $M\sim
\Lambda$ is generic (barring special cases such as Goldstone or pseudo-Goldstone
excitations).  It is then convenient to separate the rest mass from soft
departures from uniform motion by writing
\begin{equation}
p^m = M v^m + k^m,
\qquad
v^2=1,
\label{eq:residual-momentum}
\end{equation}
where $v^m$ is chosen close to the particle’s actual four-velocity (for a narrow
wavepacket one may take $v^m\simeq p^m/M$).  The residual momentum $k^m$ then
captures slow variations induced, for instance, by weak external fields, rather
than the overall laboratory momentum.

In the Brookhaven bubble–chamber observation, the $\Omega^{-}$ was produced with
$|\vec p|\simeq2~\mathrm{GeV}$ and mass $M_{\Omega^-}\simeq1.672~\mathrm{GeV}$,
corresponding to a moderate Lorentz factor $\gamma\simeq1.20$.  The chamber
operated in a magnetic field of order $B\simeq1$–$2~\mathrm{T}$, for which
\[
eB \simeq 1.95\times10^{-16}\,\mathrm{GeV}^2
\qquad (B=1~\mathrm{T}),
\]
and therefore
\[
\epsilon_{F}
= \frac{|eB|}{M_{\Omega^-}^{2}}
\sim 10^{-16}.
\]
Electromagnetic effects are thus suppressed by many orders of magnitude compared
to the hadronic scale.

The $\Omega^{-}$ also provides a striking example of a composite spin-$3/2$
state with $g\neq2$.  Its magnetic moment
$\mu_{\Omega^-}=(-2.02\pm0.05)\,\mu_N$, together with
\[
\mu = g\,\frac{Q}{2M}\,S,
\qquad
Q=-e,\ \ S=\tfrac{3}{2},
\]
yields
\[
g_{\Omega^-} \simeq 2.40\pm0.06,
\]
significantly above the pointlike value.

To estimate the residual momentum, we choose $v^m$ aligned with the observed
average motion so that $k^m$ tracks the slow bending of the trajectory.
Over a path length $L\sim2.5\,$cm, the magnetic field induces a momentum change
$\Delta p\sim |eB|\,L$.  Using
$L\simeq1.27\times10^{14}\ \mathrm{GeV}^{-1}$ gives
\[
\Delta p \sim 10^{-2}\ \mathrm{GeV},
\qquad
\epsilon_p \sim \frac{\Delta p}{M_{\Omega^-}} \sim 10^{-2}\ll1.
\]
The simultaneous smallness of $\epsilon_p$ and $\epsilon_F$ confirms that, in
this classic event, the EFT expansion in residual derivatives and local
multipole operators is fully justified.

\medskip
Our main results can be summarized as follows:
\begin{itemize}
\item Assuming the tracelessness constraint holds exactly, we show that the
pointlike spin-2 and spin-$\tfrac32$ systems in constant electromagnetic
backgrounds admit consistent constraint chains only at the universal value
$g=2$. This uniqueness is already known (and, for spin-$\tfrac32$, has been
proven for the most general form of the equations of motion
in~\cite{Benakli:2024aes}), but it provides a clean illustration of our central
diagnostic: the constraint sector remains non-degenerate
(\emph{constraint-sector closure}), so that non-physical components are fixed
rather than requiring independent initial data.

\item We construct a controlled EFT deformation that allows $g\neq 2$ in the
weak-field/near-shell regime by introducing specific off-shell completion
operators proportional to $F\,\Delta/M^2$, where $\Delta$ denotes
the free near-on-shell operator.

\item For spin-2, the deviation from transversality is \emph{induced} for
$g\neq 2$ and determined perturbatively by the divergence equation, so no extra
degrees of freedom propagate in the EFT domain: the non-transverse components
are fixed functionals of the physical modes.

\item For spin-$\tfrac32$, we provide a projector formulation in which the
$\sigma$-trace (equivalently, $\gamma$-trace) constraints are preserved
manifestly and the resulting second-order wave operator exhibits a single
physical gyromagnetic ratio multiplying both the spinor and vector Zeeman
couplings, as required for an irreducible massive spin-$\tfrac32$ multiplet.

\item In both spin-2 and spin-$\tfrac32$ cases, we illustrate the formalism by
solving the simplest situation of a constant magnetic field in Landau gauge.
We extract the leading Landau and Zeeman structure and comment on the principal
symbol and hyperbolicity of the $\mathcal O(F)$-truncated systems in this
background.
\end{itemize}

The paper is organized as follows. In Section~\ref{sec:spin2-minimal}, we review
the pointlike spin-2 system and explicitly identify the algebraic divergence
obstruction that restricts the pointlike theory to $g=2$. Building on this,
Section~\ref{sec:spin2-EFT} constructs the leading spin-2 EFT deformation and
derives the induced non-transversality condition required for consistency when
$g \neq 2$. In Section~\ref{sec:spin2-Landau}, we illustrate the resulting
equations and constraints by analysing the Landau-level structure for the
spin-2 field in a constant magnetic field. Following the bosonic analysis,
Section~\ref{sec:spin32_g2} reviews the consistent $g=2$ spin-$\tfrac{3}{2}$
system, and Section~\ref{sec:spin32_generic_EFT} presents the construction of
the corresponding generic-$g$ spin-$\tfrac{3}{2}$ EFT, including the
$F\,\Delta/M^2$ completion terms. Section~\ref{sec:spin32_landau} derives the
Landau spectrum for spin-$\tfrac{3}{2}$ in Landau gauge and discusses the
domain of validity of the EFT, and Section~\ref{sec:conclusion} summarizes our
results. Several technical derivations are collected in the appendices.

\section{Spin--2: point-like systems and the divergence obstruction}
\label{sec:spin2-minimal}

\subsection{Free Fierz--Pauli system}

A free massive neutral spin--2 particle in four dimensions may be described, at the level of equations
of motion, by a real symmetric tensor field $\Phi_{mn}=\Phi_{nm}$ obeying the Fierz--Pauli (FP)
system \cite{Fierz:1939ix}
\begin{align}
(\partial^2+M^2)\,\Phi_{mn} &= 0,
\label{eq:FP-eom-free}
\\
\Phi^m{}_m &= 0,
\qquad
\partial^m \Phi_{mn}=0.
\label{eq:FP-constraints-free}
\end{align}
The algebraic trace condition and the differential transversality condition remove the lower-spin
components contained in a generic symmetric-tensor representation (spin–1 and spin–0). Taken together,
\eqref{eq:FP-eom-free}--\eqref{eq:FP-constraints-free} propagate $2s+1=5$ physical degrees of freedom
and define a hyperbolic system.

\subsection{Charged field: covariantization and the unique dipole operator at $\mathcal O(F)$}

For a charged spin--2 field, $\Phi_{mn}$ becomes complex and derivatives are covariantized.
We consider the adjunction of a dipole-type operator of the form
\begin{equation}
E_{mn}
=
(D^2+M^2)\,\Phi_{mn}
-2ie\,\gamma\,(F\Phi-\Phi F)_{mn}
=0,
\label{eq:spin2-minimal-E-rig}
\end{equation}
where $D^2\equiv D^mD_m$ and $(F\Phi-\Phi F)_{mn}\equiv
F_m{}^{r}\Phi_{rn}-\Phi_m{}^{r}F_{rn}$. The tensor $(F\Phi-\Phi F)_{mn}$ is symmetric because for an
antisymmetric $F$ and symmetric $\Phi$ one has $(F\Phi)^T=-\Phi F$.

In a static magnetic background ($F_{0i}=0$, $F_{ij}=\epsilon_{ijk}B^k$), the operator $(F\Phi-\Phi F)_{ij}$
acts on the spatial symmetric–traceless tensor $\Phi_{ij}$ as the standard spin--2 generator of
rotations. The interaction therefore reduces to a Zeeman coupling $\propto\vec{B}\cdot\vec{S}$, and with
the conventional choice of generators one identifies the effective gyromagnetic ratio
\begin{equation}
g_{\rm eff}=2\gamma.
\label{eq:geff-gamma}
\end{equation}

\subsection{Divergence equation and closure on the constraint surface}

Define the covariant transversality quantity
\begin{equation}
J_n \equiv D^m \Phi_{mn}.
\label{eq:Jdef}
\end{equation}
A basic consistency question is whether the restriction to the physical sector, specified by the initial
conditions, here the covariantized version of the FP condition $J_n=0$ together with the trace constraint, is preserved under time evolution by the
equation of motion \eqref{eq:spin2-minimal-E-rig}.

To test this, one takes a covariant divergence of \eqref{eq:spin2-minimal-E-rig} and examines whether it
vanishes on the constraint surface. For constant $F$ ($D_\ell F_{mn}=0$) and using
\begin{equation}
[D^m,D^2]=2 i e\,F^{m}{}_{r}D^r,
\end{equation}
one obtains
\begin{align}
D^m E_{mn}
&=
(D^2+M^2)J_n
+2ie(1-\gamma)\,F_{m}{}^{r}D^m \Phi_{rn}
+2ie\,\gamma\,F_{rn}J^r.
\label{eq:div-minimal-rig}
\end{align}
The tensor $F_{m}{}^{r}D^m \Phi_{rn}$ is an independent structure and is not proportional to
$J_n$. Thus, exact closure on the constraint surface requires its coefficient to vanish:
\begin{equation}
\gamma=1
\qquad\Longleftrightarrow\qquad
g_{\rm eff}=2.
\label{eq:gamma1}
\end{equation}
Hence, for a point-like single-field system of the form \eqref{eq:spin2-minimal-E-rig}, consistency
uniquely selects the value $g_{\rm eff}=2$.
\section{Spin--2 EFT for composite particles with $g\neq 2$:
constraint-sector closure and induced non-transversality}
\label{sec:spin2-EFT}

We now describe an EFT interpretation appropriate for \emph{composite} charged massive spin--2 particles, for
which a generic magnetic moment (and hence $g\neq 2$) is expected. The relevant requirement is
\emph{constraint-sector closure}: within the EFT regime $|eF|\ll M^2$ and for small residual variations,
the non-transverse components are determined perturbatively as functionals of the physical components and
do not require additional initial data. No new degrees of freedom appear.

Again, in practice one phrases this in terms of the vanishing of the divergence of the deformed equations
of motion. For a symmetric spin--2 field in a constant background ($D_\ell F_{mn}=0$), the equations
$E_{mn}(\Phi;F)=0$ imply a relation of the schematic form
\begin{equation}
D^m E_{mn}=0
\qquad\Longrightarrow\qquad
\mathcal K_n{}^{r} J_r = \mathcal S_n[\Phi;F],
\label{eq:generic-constraint-equation-spin2}
\end{equation}
where $J_n$ was defined in \eqref{eq:Jdef}, $\mathcal K$ is a massive differential operator, and
$\mathcal S_n$ is a source built from $\Phi_{mn}$ and $F_{mn}$.

In the point-like case, consistency requires the divergence equation to be \emph{homogeneous} in $J_n$
(i.e.\ $\mathcal S_n=0$), so that $J_n=0$ is preserved exactly once imposed on the initial data; this
uniquely selects $\gamma=1$ (equivalently $g_{\rm eff}=2$).

For composite states, by contrast, one allows $\mathcal S_n\neq 0$, so that non-transversality is
\emph{induced}:
\begin{equation}
D^m\Phi_{mn} = \mathcal O\!\left(\frac{eF}{M^2}\right).
\end{equation}
Here “induced’’ means that $J_n$ is not a new free datum: it is fixed order by order in $F$ and in
residual derivatives by solving Eq.~\eqref{eq:generic-constraint-equation-spin2}. This controlled
deformation of the constraint does not introduce additional propagating degrees of freedom.

A natural first step might be to try to cancel the obstruction term $F_{m}{}^{r}D^m\Phi_{rn}$ in the
divergence of the minimal equation by adding local operators with one derivative on $\Phi$ (schematically
$\mathcal O(F\,D\Phi)$). However, genuinely \emph{small} deformations in the EFT sense do not achieve this.
Moreover, any operator of the form $\mathcal O(F\,D\Phi)$ inevitably generates, via commutators,
contributions with \emph{two} derivatives on $\Phi$ (schematically $\mathcal O(F\,D^2\Phi)$), up to
$\mathcal O(F^2)$ corrections. Such terms cannot remove the problematic $\mathcal O(FD\Phi)$ tensor
structure and may in fact introduce additional ones.

This motivates working directly with operators that contain two derivatives and are organised in terms of
insertions of the off-shell operator $\Delta\equiv D^2+M^2$, i.e.\ structures of the form
$\mathcal O(F\,\Delta\Phi/M^2)$. In the EFT regime, $\Delta$ is small on near-on-shell modes
($\Delta\Phi\sim Mk\,\Phi$), so operators proportional to $F\Delta/M^2$ are parametrically suppressed
relative to the leading dipole term and provide a natural way to deform the point-like $g=2$ system while
maintaining control over the constraint sector.

\subsection{EFT equations of motion, the redundant operator, and $g_{\rm eff}$}

We organise the EFT using $\Delta \equiv D^2+M^2$, and include, at $\mathcal O(F)$, the dipole (Pauli)
operator plus a convenient completion:
\begin{equation}
E_{mn}[\Phi;F]
=
\Delta\,\Phi_{mn}
-2ie\,\gamma\,
\bigl(F_{m}{}^{r}\Phi_{rn}-\Phi_{m}{}^{r}F_{rn}\bigr)
+\frac{ie\,\beta}{M^{2}}\,
F_{(m}{}^{r}\,\Delta \Phi_{n)r}
=0.
\label{eq:spin2-EOM-rig}
\end{equation}
(The sign in front of the Pauli term is chosen to match the point-like case
\eqref{eq:spin2-minimal-E-rig}.)

The parameter $\gamma$ fixes the physical dipole coupling. The $\beta$--term is proportional to
$\Delta\Phi$ and hence can be neglected when the mass-shell condition $\Delta\Phi_{mn}=\mathcal O(eF)\,\Phi$
is imposed. Accordingly, at $\mathcal O(F)$ the on-shell magnetic moment is controlled only by $\gamma$,
\begin{equation}
g_{\rm eff}=2\gamma.
\end{equation}

The coefficient $\beta$ multiplies a \emph{redundant} operator: it is proportional to the leading
equations of motion and can be shifted (or set to zero) by a local field redefinition without changing
on-shell observables at the working order. Nevertheless, we choose to keep $\beta$ explicit as it is
useful to track how the \emph{derived constraint relations} are modified away from the free mass shell.

\subsection{Divergence and the induced constraint (subsidiary) equation}
\label{sec:spin2-induced-nontransversality}

Taking a covariant divergence of the EFT equations of motion \eqref{eq:spin2-EOM-rig} and working to
$\mathcal O(F)$ gives
\begin{align}
D^mE_{mn}
&=
\Delta J_n
+2ie(1-\gamma)\,F_{m}{}^{r}D^m\Phi_{rn}
+2ie\gamma\,F_{rn}J^r
\nonumber\\
&\hspace{1.2cm}
+\frac{ie\,\beta}{2M^2}\Bigl[
F_{m}{}^{r}\,\Delta(D^m\Phi_{nr})
+F_{n}{}^{r}\,\Delta J_r
\Bigr]
+\mathcal O(F^2).
\label{eq:div-EFT-rig}
\end{align}
Here we have used $D_\ell F_{mn}=0$ and $[D^m,\Delta]=2ie\,F^{m}{}_{r}D^r$, so the commutator acting on
the $\beta$–term produces an extra factor of $F$ and is therefore $\mathcal O(F^2)$, which we
systematically discard in this section.

Equation \eqref{eq:div-EFT-rig} is the \emph{subsidiary (constraint-propagation) equation} associated
with the divergence of the spin--2 equations. In the special point-like case $\gamma=1$ (equivalently
$g_{\rm eff}=2$) the source term proportional to $F_m{}^{r}D^m\Phi_{rn}$ drops out and the relation is
homogeneous in $J$, so the constraint surface $J_n=0$ is preserved.

For generic $\gamma\neq 1$, the tensor structure $F_m{}^{r}D^m\Phi_{rn}$ is independent and not
proportional to $J_n$, so \eqref{eq:div-EFT-rig} becomes inhomogeneous. We interpret this not as an
additional independent field equation, but as a derived relation that determines the non-transverse
components (encoded in $J_n$) perturbatively in the EFT regime.

\subsection{Determination of $J_n$ and EFT closure}

The divergence relation \eqref{eq:div-EFT-rig} is an equation for the derived quantity
$J_n \equiv D^m\Phi_{mn}$. Since in the EFT $J_n$ is not an independent field variable, solving
\eqref{eq:div-EFT-rig} does not introduce new degrees of freedom; rather, it provides (together with the
trace constraint) the relations needed to eliminate non-physical components in terms of the physical
spin--2 sector.

It is useful to rewrite \eqref{eq:div-EFT-rig} schematically as
\begin{equation}
\mathcal K_{n}{}^{r}\,J_r \;=\; \mathcal S^{(1)}_n[\Phi;F]\;+\;\mathcal O(F^2),
\label{eq:spin2_KJ_S_form}
\end{equation}
where, to $\mathcal O(F)$,
\begin{align}
\mathcal K_{n}{}^{r}
&=
\Delta\,\delta_{n}^{\ r}
+2ie\gamma\,F_{n}{}^{r}
+\frac{ie\beta}{2M^2}\,F_{n}{}^{r}\,\Delta,
\\
\mathcal S^{(1)}_n[\Phi;F]
&=
-2ie(1-\gamma)\,F_{m}{}^{r}D^m\Phi_{rn}
-\frac{ie\beta}{2M^2}\,F_{m}{}^{r}\,\Delta(D^m\Phi_{nr}).
\end{align}
EFT \emph{constraint-sector closure} means that, in the weak-field/near-shell regime, $\mathcal K$ is
invertible order by order in $(\epsilon_F,\epsilon_\Delta)$ on the modes of interest, so that $J_n$ is
fixed as a functional of $\Phi$ rather than requiring independent initial data.

Formally one may write $J=\mathcal K^{-1}\mathcal S$. This is an operator inversion and is not
automatically local. In the EFT regime one has two equivalent practical implementations:

(i) a \emph{derivative} expansion of $\mathcal K^{-1}$ in powers of $\Delta/M^2$ and $eF/M^2$ (yielding a
local series to the working order), or

(ii) a \emph{mode-by-mode} solution (e.g.\ in the Landau basis), where $\Delta$ becomes an eigenvalue and
\eqref{eq:spin2_KJ_S_form} reduces to a finite-dimensional linear system in Lorentz indices.

At the lowest order in the combined $(eF/M^2,\Delta/M^2)$ counting, one may drop the $\beta$-suppressed
terms and treat the inversion of $\mathcal K$ perturbatively in $eF/M^2$, which gives the schematic leading
behaviour
\begin{equation}
J_n
=
-\,\frac{2ie(1-\gamma)}{M^2}\,F_{m}{}^{r}D^m\Phi_{rn}
\;+\;\mathcal O\!\left(F\,\frac{\Delta}{M^2}\Phi\right)
\;+\;\mathcal O(F^2).
\label{eq:J_leading_EFT_estimate}
\end{equation}
Equation~\eqref{eq:J_leading_EFT_estimate} is simply the first term of the EFT
expansion of $\mathcal K^{-1}$: since the non-physical components are
off-shell by~$\sim M$, inverting $\mathcal K$ reduces at leading order to
dividing by $M^2$.  All further terms are suppressed by $\Delta/M^2$ or by an
additional power of $F$.

In the constant magnetic field analysis of Sec.~\ref{sec:spin2-Landau}, one may instead work directly
mode by mode. Neglecting $\beta$-suppressed contributions at leading order in residual derivatives, one
obtains on each Landau mode
\begin{equation}
\left[ \Delta(E,p_z,\ell)\,\delta_n^{\ r} - 2ie\gamma\, F_n{}^{r} \right] J_r
=
-2ie(1-\gamma)\,\bigl(F_m{}^{s}D^m\Phi_{s n}\bigr)
\;+\;\mathcal O(F^2),
\label{eq:J-mode-equation}
\end{equation}
which makes the meaning of the formal $\mathcal K^{-1}$ explicit: it is simply division by the shifted
Landau eigenvalue together with inversion of the finite-dimensional Lorentz-index mixing induced by
$F_n{}^{r}$.


Two important remarks are useful at this point:

\begin{itemize}

\item \textbf{Constraint-sector closure and degrees of freedom.}
Within the EFT regime characterized by $|eF| \ll M^2$ and the truncation implicit in
\eqref{eq:spin2-EOM-rig}, the divergence relation \eqref{eq:div-EFT-rig} determines $J_n$ perturbatively
(either via a local derivative expansion or mode by mode as in \eqref{eq:J-mode-equation}).
Consequently, we stress again that the constraint sector closes order by order: the non-transverse components are fixed by the
equations (and the trace constraint) and do not require independent initial data. This is essential in
avoiding additional degrees of freedom. The redundant $\beta$-operator is proportional to $\Delta\Phi$ and
does not affect the $\mathcal O(F)$ dipole matching $g_{\rm eff}=2\gamma$.

\item \textbf{Distinction from global hyperbolicity}
This perturbative closure statement is obviously not a global proof of causal propagation for arbitrary field
strengths. The Velo--Zwanziger issue concerns the coupled constrained PDE system and the possible
degeneration of the constraint chain (equivalently, loss of rank in the subsidiary relations) outside the
weak-field regime. Here we restrict to the EFT domain, where \eqref{eq:div-EFT-rig} provides a controlled
determination of the non-transverse sector consistent with the spectral analysis.

\end{itemize}

\section{Motion of a spin--2 particle in a constant magnetic field}
\label{sec:spin2-Landau}

We now apply the general EFT framework to the simplest non-trivial background: a constant magnetic field $\vec{B}=B\hat{z}$. This explicit calculation demonstrates 
how the effective gyromagnetic ratio $g_{\rm eff}$ controls the Zeeman splitting and how the induced non-transversality is handled perturbatively.

In the relevant regime --- weak background fields and modes near the one-particle mass shell --- the expansion is controlled by the small parameters
\begin{equation}
\frac{|eF|}{M^2}\ll 1,
\qquad
\left|\frac{\Delta}{M^2}\right|\ll 1.
\label{eq:EFT-counting-Landau}
\end{equation}
As shown below, in a constant-$B$ background the eigenvalues of the operator $\Delta \equiv D^2+M^2$ scale as $|eB|/M^2$ for low-lying Landau levels. 
Consequently, the term proportional to $\beta$ in Eq.~\eqref{eq:spin2-EOM-rig}, which involves the operator $F\Delta\Phi$, is suppressed by an additional 
factor of $\mathcal{O}(|eF|/M^2)$ relative to the leading Pauli coupling. To linear order in the EFT expansion, we may therefore neglect the $\beta$-term. 
The equations of motion reduce to
\begin{equation}
\bigl(D^{2} + M^{2}\bigr)\Phi_{mn}
-2ie\,\gamma\,
\bigl(F_{m}{}^{r}\Phi_{rn}-\Phi_{m}{}^{r}F_{rn}\bigr)=0,
\label{eq:EOM-Landau-effective}
\end{equation}
where the effective gyromagnetic ratio is identified as $g_{\rm eff}=2\gamma$.

\subsection{Landau levels and induced non-transversality in practice}
%

We consider a static, homogeneous magnetic field along the $z$-axis ($F_{12}=B$, $F_{0m}=F_{3m}=0$) and adopt the Landau gauge $A_m=(0,\,0,\,Bx,\,0)$. The covariant derivatives for a field of charge $e$ are
\begin{align}
D_0&=\partial_t, &
D_1&=\partial_x, &
D_2&=\partial_y+ieBx, &
D_3&=\partial_z,
\end{align}
satisfying $[D_1,D_2]= ieB$. The covariant d'Alembertian in the $(+---)$ signature decomposes as
\begin{equation}
D^{2} = \partial_t^2 - \partial_x^2 - (\partial_y + i e B x)^2 - \partial_z^2.
\label{eq:D2-Landau}
\end{equation}
We separate variables using the ansatz
\begin{equation}
\Phi_{mn}(t,x,y,z) = e^{-iEt+ip_yy+ip_zz}\,\phi_{mn}(x).
\label{eq:plane-wave-ansatz-Landau}
\end{equation}
Acting on this mode, the wave operator reduces to
\begin{equation}
D^2 \Phi_{mn}
=
e^{-iEt+ip_yy+ip_zz}\,
\Bigl[-E^2 + p_z^2 - \partial_x^2 + (p_y + eBx)^2\Bigr]\phi_{mn}(x).
\label{eq:D2-on-phi}
\end{equation}
The transverse part is recognized as the Hamiltonian of a shifted harmonic oscillator,
$H_B \equiv -\partial_x^2 + \pi_y(x)^2$, with $\pi_y(x) \equiv p_y + eBx$.


Before diagonalizing the spectrum, we address the constraint sector.
In the point-like limit ($g_{\rm eff}=2$), the transversality condition $J_n \equiv D^m \Phi_{mn} = 0$ holds exactly. This allows the non-dynamical components $\phi_{0m}$ to be eliminated algebraically in terms of the spatial components $\phi_{ij}$.

For composite states with $g_{\rm eff}\neq 2$, the quantity $J_n$ is no longer zero but is determined perturbatively by the equations of motion. The leading EFT relation derived in Sec.~\ref{sec:spin2-EFT} is
\begin{equation}
J_n = -\frac{2ie(1-\gamma)}{M^2}\,F_m{}^{r}D^m\Phi_{rn} + \mathcal{O}(F^2).
\label{eq:J-induced-curl}
\end{equation}
In the specific background considered here, this implies that the $n=0$ component of the constraint becomes an inhomogeneous relation,
\begin{equation}
-iE\,\phi_{00} - \partial_x\phi_{01} - i\pi_y\,\phi_{02} - ip_z\,\phi_{03} \,=\, J_0,
\label{eq:div-n0-induced}
\end{equation}
where the source term $J_0$ is induced by spatial gradients of the physical fields:
\begin{equation}
J_0 \simeq -\frac{2ie(1-\gamma)B}{M^2} \bigl(\partial_x\phi_{20}-i\pi_y\,\phi_{10}\bigr).
\end{equation}
Eq.~\eqref{eq:div-n0-induced} can be solved for $\phi_{00}$ (and similarly for $\phi_{0i}$), determining the non-transverse components as functionals of the spatial components to the desired order in $\mathcal{O}(eB/M^2)$. The algebraic trace constraint $\Phi^m{}_m=0$ remains unmodified at this order. Consequently, the number of propagating degrees of freedom remains five, despite the deformation of the constraint surface.


We now determine the spectrum using perturbation theory. At leading order, we solve the equations in the transverse-traceless (TT) sector, treating the mixings induced by $J_n \neq 0$ as higher-order EFT corrections.

The TT components are conveniently packaged in the helicity basis diagonalizing $S_z$:
\begin{align}
\Phi^{(\pm 2)} &= \tfrac12(\Phi_{11}-\Phi_{22}\mp2i\,\Phi_{12}), &
\Phi^{(\pm 1)} &= \tfrac{1}{\sqrt2}(\Phi_{13}\mp i\,\Phi_{23}), &
\Phi^{(0)} &= \tfrac{1}{\sqrt6}(2\Phi_{33}-\Phi_{11}-\Phi_{22}).
\end{align}
In this basis, the Pauli interaction term in Eq.~\eqref{eq:EOM-Landau-effective} reduces to the diagonal Zeeman operator $-2e\gamma B \lambda$. The leading wave equation for the TT modes is then
\begin{equation}
\Bigl[ D^2 + M^2 - 2e\gamma B \lambda \Bigr] \Phi^{(\lambda)} = 0, \qquad \lambda \in \{0, \pm 1, \pm 2\}.
\end{equation}

Substituting the eigenvalues of the transverse oscillator, $H_B \to |eB|(2\ell+1)$, we obtain the dispersion relation
\begin{equation}
E_{\ell,\lambda}^2
=
M^2 + p_z^2 + |eB|(2\ell +1)
- g_{\rm eff}\, eB\,\lambda
\;+\; \delta E_{\ell,\lambda}^2,
\label{eq:Landau-spectrum-spin2-leading}
\end{equation}
where $\ell=0,1,2,\dots$ is the Landau level index.
The correction term $\delta E_{\ell,\lambda}^2$ arises from the induced non-transversality $J_n \neq 0$. Since $J_n$ scales as $\mathcal{O}(|eB|/M^2)$ and involves ladder operators (via $D_{1,2}$), it induces mixing between neighbouring levels $\ell \to \ell \pm 1$. These effects contribute to the energy squared at order $\mathcal{O}(|eB|^2/M^2)$, which is consistent with the truncation error of the EFT.


The validity of the local EFT expansion relies on a hierarchy of scales. In the Landau problem, the transverse covariant derivatives scale as $D_\perp \sim \sqrt{|eB|(2\ell+1)}$. Therefore, the expansion of the inverse operator $\Delta^{-1}$ (required to solve for $J_n$) corresponds to an asymptotic series in the parameters
\begin{equation}
\epsilon_B = \frac{|eB|}{M^2} \ll 1, \qquad
\epsilon_\ell = \frac{|eB|\ell}{M^2} \ll 1.
\end{equation}
The condition $\epsilon_\ell \ll 1$ implies that the EFT description holds for low-lying Landau levels. For highly excited states where $\ell \sim M^2/|eB|$, the residual momentum becomes comparable to the mass, and the local derivative expansion breaks down.

\subsection{Causality and hyperbolicity in the magnetic background}
\label{sec:spin2-hyperbolicity}

The Landau-level spectral analysis presupposes that the underlying field equations describe a well-posed initial value problem. We now explicitly check the causal structure of the EFT equations in the constant magnetic background to verify that no Velo-Zwanziger pathologies (such as acausal propagation or loss of hyperbolicity) arise at the working order.

We analyze the Cauchy problem for the truncated $\mathcal{O}(F)$ system with respect to the standard time $t=x^0$:
\begin{equation}
E_{mn}[\Phi;F]
=
\Delta\,\Phi_{mn}
-2ie\,\gamma\,
\bigl(F_{m}{}^{r}\Phi_{rn}-\Phi_{m}{}^{r}F_{rn}\bigr)
+\frac{ie\,\beta}{M^{2}}\,
F_{(m}{}^{r}\,\Delta \Phi_{n)r}
=0.
\label{eq:spin2-EOM-hyp}
\end{equation}
In the static magnetic background ($F_{12}=B$), we choose the gauge $A_0=0$, so $D_0=\partial_t$ and $[D_0,D_i]=0$.

Hyperbolicity is determined by the highest-derivative operators in the equation.
The Pauli term (proportional to $\gamma$) is algebraic in derivatives and does not contribute to the principal symbol. The $\beta$-term involves two derivatives (via $\Delta$), but in a constant background it enters as a spacetime-independent linear map acting on the indices of the wave operator. We may factor the equation as:
\begin{equation}
E_{mn} = (\mathsf C\,\Delta\Phi)_{mn} + \text{(lower-derivative terms)},
\end{equation}
where the endomorphism $\mathsf C$ acting on symmetric tensors is defined by
\begin{equation}
(\mathsf C X)_{mn} \equiv X_{mn}+\frac{ie\beta}{2M^2}\big(F_m{}^{r}X_{rn}+F_n{}^{r}X_{rm}\big).
\end{equation}
In the EFT regime $|eF|/M^2 \ll 1$, the map $\mathsf C$ is a small perturbation of the identity, $\mathsf C = \mathbf{1} + \mathcal{O}(\epsilon_F)$, and is therefore invertible. Multiplying the equations from the left by $\mathsf C^{-1}$ isolates the wave operator $\Delta \Phi$ as the principal part. Consequently, the characteristic surfaces are determined solely by the principal symbol of $\Delta = D^2 + M^2$, which coincides with the light cone of the background metric.

To demonstrate strong hyperbolicity explicitly, we perform a first-order reduction (details are given in Appendix~\ref{app:spin2-hyperbolicity}). We introduce the variables
\begin{equation}
\Pi_{mn}\equiv \partial_t\Phi_{mn},
\qquad
Q_{i\,|\,mn}\equiv D_i\Phi_{mn} \quad (i=1,2,3),
\end{equation}
and collect them into the state vector $U\equiv(\Phi_{mn},\Pi_{mn},Q_{i\,|\,mn})$.
Using the $3+1$ decomposition of the wave operator $D^2 = \partial_t^2-\delta^{ij}D_iD_j$ (where $D_i$ can be replaced by $\partial_i$ in the principal part), the system reduces to:
\begin{equation}
\partial_t\Phi_{mn}=\Pi_{mn},\qquad
\partial_t Q_{i\,|\,mn}=\partial_i\Pi_{mn},\qquad
\partial_t\Pi_{mn}=\delta^{jk}\partial_j Q_{k\,|\,mn}.
\label{eq:PiQ-principal-short-fixed}
\end{equation}
For any spatial wavevector $k_i$, the principal symbol matrix $\mathbb A(k)$ in the $(\Pi,Q)$ block takes the standard form:
\begin{equation}
\mathbb A(k)\Big|_{(\Pi,Q)}=
\begin{pmatrix}
0 & k^i\\
k_i & 0
\end{pmatrix},
\quad \text{where } k^i \equiv \delta^{ij}k_j.
\label{eq:Aprincipal-short-fixed}
\end{equation}
The eigenvalues are $\omega = \pm |k|$, which are real and distinct. Since the symbol is uniformly diagonalizable, the system is strongly hyperbolic.

We conclude that in a constant magnetic background, the $\mathcal{O}(F)$-truncated EFT system does not suffer from the Velo-Zwanziger instability. The characteristic speeds remain luminal, and the Cauchy problem is well-posed. The non-minimal couplings ($\gamma$ and $\beta$) modify the system either at the lower-derivative level or by an invertible field redefinition ($\mathsf C$), neither of which destroys the hyperbolicity of the principal wave operator in the weak-field regime.

\section{Spin-$\tfrac{3}{2}$: point-like systems and the $g=2$ value}
\label{sec:spin32_g2}

The field content is a left-handed vector--spinor $\chi_m$ and a right-handed partner $\bar\lambda_m$
(the bar denotes a dotted spinor index). For a charged field, $\chi_m$ and $\bar\lambda_m$ are treated as
independent complex fields of charge $e$.

\subsection{First-order system and algebraic constraints}

A convenient first-order formulation of the system is~\cite{Benakli:2023aes,Benakli:2022edf}

\begin{align}
i\,\bar\sigma^n D_n \chi_m + M\,\bar\lambda_m &= 0,
\label{eq:RS1}
\\
i\,\sigma^n D_n \bar\lambda_m + M\,\chi_m &= -\,i\frac{e\,\zeta}{M}\,F_{mn}\chi^n ,
\label{eq:RS2zeta}
\end{align}
supplemented by the algebraic $\sigma$-trace constraints
\begin{equation}
\bar\sigma^m\chi_m = 0,
\qquad
\sigma^m\bar\lambda_m = 0.
\label{eq:sigmatrace32}
\end{equation}
The parameter $\zeta$ controls the non-minimal coupling associated with the \emph{vector} index of the
spin-$\tfrac{3}{2}$ field. We now show that closure of the constraint chain fixes
\begin{equation}
\zeta=2\qquad\Longleftrightarrow\qquad g=2.
\end{equation}

\subsection{Secondary (divergence) constraints}

Contract \eqref{eq:RS1} with $\sigma^m$:
\begin{equation}
\sigma^m\big(i\bar\sigma^n D_n\chi_m\big)+M\,\sigma^m\bar\lambda_m=0.
\end{equation}
Using $\sigma^m\bar\sigma^n=2\eta^{mn}-\sigma^n\bar\sigma^m$ and the algebraic constraint
$\sigma^m\bar\lambda_m=0$ gives
\begin{equation}
2i\,D^m\chi_m - i\,\sigma^n D_n(\bar\sigma^m\chi_m)=0,
\end{equation}
and with $\bar\sigma^m\chi_m=0$ this reduces to
\begin{equation}
D^m\chi_m=0.
\label{eq:divchi32}
\end{equation}

Next, contract \eqref{eq:RS2zeta} with $\bar\sigma^m$:
\begin{equation}
\bar\sigma^m\big(i\sigma^n D_n\bar\lambda_m\big)+M\,\bar\sigma^m\chi_m
=
-\,i\frac{e\zeta}{M}\,\bar\sigma^mF_{mn}\chi^n.
\end{equation}
Using $\bar\sigma^m\sigma^n=2\eta^{mn}-\bar\sigma^n\sigma^m$ together with
$\bar\sigma^m\chi_m=0$ and $\sigma^m\bar\lambda_m=0$, the left-hand side becomes $2iD^m\bar\lambda_m$,
so that
\begin{equation}
D^m\bar\lambda_m
=
-\frac{e\zeta}{2M}\,\bar\sigma^m F_{mn}\chi^n.
\label{eq:divlambda32_algebraic}
\end{equation}

\subsection{Constraint preservation and the divergence obstruction}
\label{sec:spin32_div_obstruction}

As for the spin--2 case, the consistency of the system is probed by evaluating the divergence of the equations of motion.
Apply $D^m$ to \eqref{eq:RS1}:
\begin{equation}
i\,\bar\sigma^n D^mD_n\chi_m + M\,D^m\bar\lambda_m=0.
\end{equation}
to get
\begin{equation}
-\,e\,\bar\sigma^nF^{m}{}_{n}\chi_m + M\,D^m\bar\lambda_m=0
\quad\Longrightarrow\quad
D^m\bar\lambda_m
=
-\frac{e}{M}\,\bar\sigma^mF_{mn}\chi^n.
\label{eq:divlambda32_dynamical}
\end{equation}
Comparing \eqref{eq:divlambda32_dynamical} with the algebraic relation
\eqref{eq:divlambda32_algebraic} immediately yields
\begin{equation}
\zeta=2.
\label{eq:zeta2_from_div32}
\end{equation}
Thus closure of the constraint chain for the point-like system fixes the
vector-index magnetic coupling to the value corresponding to $g=2$.

\subsection{Second-order wave equations and magnetic moment}

Define
\begin{equation}
\sigma^{mn}\equiv \frac{i}{4}(\sigma^m\bar\sigma^n-\sigma^n\bar\sigma^m),
\qquad
\bar\sigma^{mn}\equiv \frac{i}{4}(\bar\sigma^m\sigma^n-\bar\sigma^n\sigma^m).
\end{equation}
With our conventions one has the Weyl-squaring identities
\begin{equation}
(\sigma\!\cdot\! D)(\bar\sigma\!\cdot\! D)= D^2 + e\,\sigma^{mn}F_{mn},
\qquad
(\bar\sigma\!\cdot\! D)(\sigma\!\cdot\! D)= D^2 + e\,\bar\sigma^{mn}F_{mn}.
\label{eq:Weyl-square32}
\end{equation}
.

Eliminate $\bar\lambda_m$ using \eqref{eq:RS1}:
\begin{equation}
\bar\lambda_m=-\frac{i}{M}\,\bar\sigma^kD_k\chi_m,
\end{equation}
insert into \eqref{eq:RS2zeta}, and multiply by $M$ to obtain
\begin{equation}
(\sigma\!\cdot\! D)(\bar\sigma\!\cdot\! D)\chi_m + M^2\chi_m
=
-\,i e\zeta\,F_{mn}\chi^n.
\end{equation}
Using \eqref{eq:Weyl-square32}, this becomes
\begin{equation}
(D^2+M^2)\chi_m + e\,\sigma^{rs}F_{rs}\,\chi_m + i e\,\zeta\,F_{mn}\chi^n = 0.
\label{eq:chi-secondorder32}
\end{equation}

Similarly, act with $i\,\bar\sigma\!\cdot\!D$ on \eqref{eq:RS2zeta}. Using $D_\ell F_{mn}=0$ and
\eqref{eq:RS1} to replace $i\bar\sigma\!\cdot\!D\,\chi_m=-M\bar\lambda_m$, one finds
\begin{equation}
(D^2+M^2)\bar\lambda_m + e\,\bar\sigma^{rs}F_{rs}\,\bar\lambda_m + i e\,\zeta\,F_{mn}\bar\lambda^n = 0.
\label{eq:lambda-secondorder32}
\end{equation}

In a purely magnetic background ($F_{0i}=0$, $F_{ij}=\epsilon_{ijk}B^k$), the term
$e\,\sigma^{rs}F_{rs}$ produces the standard spin-$\tfrac12$ Zeeman coupling on the spinor index, while
the term $i e\zeta\,F_{mn}(\cdot)^n$ produces the Zeeman coupling on the vector index (since $-iF$ is the
spin--1 generator). At $\zeta=2$ these two contributions have the same normalization and combine into a
single interaction proportional to $\vec B\!\cdot\!\vec S$ for the physical spin-$\tfrac32$ multiplet,
corresponding to the universal gyromagnetic ratio $g=2$.

\section{Spin-$\tfrac{3}{2}$ with generic gyromagnetic ratio: EFT deformation at $\mathcal O(F)$}
\label{sec:spin32_generic_EFT}

We now describe a massive charged spin-$\tfrac{3}{2}$ field with a generic gyromagnetic ratio $g\neq 2$ as an
effective field theory (EFT) in a weak, constant electromagnetic background. As in the spin--2 case, we introduce
the near-shell operator
\begin{equation}
\Delta \equiv D^2+M^2 .
\label{eq:Delta0_def_generic}
\end{equation}

We work perturbatively in the weak-field / near-shell regime
\begin{equation}
\epsilon_F\equiv \frac{|eF|}{M^2}\ll 1,
\qquad
\epsilon_\Delta\equiv \frac{\|\Delta\|}{M^2}\ll 1
\quad \text{on the modes of interest}.
\label{eq:powercount32_generic}
\end{equation}
In particular, in a constant magnetic background one has $D_\perp\sim\sqrt{|eB|}$, so the derivative
expansion is controlled by $|eB|/M^2\ll 1$. The relevant consistency requirement is again
\emph{constraint-sector closure}: within this regime, the non-physical components are fixed
perturbatively as functionals of the physical spin-$\tfrac32$ degrees of freedom and do not require
additional initial data.

\subsection{Geometric projector deformation (manifest spin-$\tfrac32$ EFT)}
\label{sec:spin32_projector_EFT}

A convenient way to describe $g\neq 2$ while maintaining control over the spin-$\tfrac32$ constraint structure
is to formulate the EFT directly in the projected spin-$\tfrac32$ subspace. This avoids unphysical
mixing with the spin-$\tfrac12$ sector and makes preservation of the $\sigma$-trace constraints
manifest.

We introduce the standard Weyl-space projector
\begin{equation}
\mathbb P_m{}^{n}\equiv\delta_m^{\,n}-\frac14\,\sigma_m \bar\sigma^{\,n},
\qquad
\bar\sigma^m\,\mathbb P_m{}^{n}=0,
\qquad
\mathbb P_m{}^{r}\mathbb P_r{}^{n}=\mathbb P_m{}^{n}.
\label{eq:P32_def}
\end{equation}

\subsection{First-order system and projected deformation}

We keep the first equation minimal (this fixes the relation between the two Weyl partners) and deform
the second equation by projected operators linear in $F$. We adopt the same $g=2$ sign convention
as in Sec.~\ref{sec:spin32_g2}:
\begin{align}
\mathcal E^{(1)}_m &\equiv
i\,\bar\sigma^n D_n \chi_m + M\,\bar\lambda_m = 0,
\label{eq:E1_proj}
\\
\mathcal E^{(2)}_m &\equiv
i\,\sigma^n D_n \bar\lambda_m + M\,\chi_m
+ i\frac{2e}{M}\,F_{mn}\chi^n
+ i\frac{e}{M}\,\mathbb P_m{}^{r}\,\mathcal K_r
=0.
\label{eq:E2_proj}
\end{align}
Equivalently,
\[
i\sigma\!\cdot\!D\,\bar\lambda_m + M\chi_m
= -\,i\frac{2e}{M}F_{mn}\chi^n - i\frac{e}{M}\mathbb P_m{}^r\mathcal K_r.
\]

At linear order in $F$ and up to dimension~7, a convenient local basis for $\mathcal K_r$ that yields a
\emph{single physical $g$ multiplying both Zeeman structures} in the squared equation is
\begin{equation}
\mathcal K_r
=
(g-2)\,F_{rn}\chi^n
-\; i\,\frac{(g-2)}{2}\,(\sigma^{ab}F_{ab})\,\chi_r
+\frac{c_{\Delta,3/2}}{M^2}\,F_{rn}\Delta\chi^n
+\cdots,
\label{eq:K_basis_fixed}
\end{equation}
where $c_{\Delta,3/2}$ is an EFT Wilson coefficient multiplying the off-shell operator $F\Delta\chi$
(suppressed by $\epsilon_\Delta$ on near-shell modes). The ellipsis denotes higher-order operators in the combined
$(\epsilon_F,\epsilon_\Delta)$ expansion.

One may ask why the deformation is placed entirely in $\mathcal E^{(2)}_m$.
A generic Pauli-type deformation of $\mathcal E^{(1)}_m$ would spoil the $\sigma$-trace derivation of
the transversality constraint, because in general $\sigma^m(\text{Pauli})_m\neq0$.
In an EFT, operators related by lower-order equations of motion are redundant, so one can choose a basis
in which the entire linear-in-$F$ deformation is assigned to $\mathcal E^{(2)}_m$ and then projected by $\mathbb P$.
This choice makes the constraint algebra manifest and is particularly well suited to the spin-$\tfrac32$ case.

\subsection{Automatic preservation of the $\sigma$-trace constraints}

We impose the algebraic $\sigma$-trace constraints
\begin{equation}
\bar\sigma^m\chi_m=0,
\qquad
\sigma^m\bar\lambda_m=0.
\label{eq:sigmatrace_proj}
\end{equation}

Contract \eqref{eq:E2_proj} with $\bar\sigma^m$. Using
$\bar\sigma^m\sigma^n=2\eta^{mn}-\bar\sigma^n\sigma^m$
and $\sigma^m\bar\lambda_m=0$, the kinetic term gives $2iD^m\bar\lambda_m$, while the mass term vanishes
because $\bar\sigma^m\chi_m=0$. The projected deformation drops out identically thanks to
$\bar\sigma^m\mathbb P_m{}^{r}=0$. Therefore,
\begin{equation}
2i\,D^m\bar\lambda_m
=
-\,i\frac{2e}{M}\,\bar\sigma^mF_{mn}\chi^n
\qquad\Longrightarrow\qquad
D^m\bar\lambda_m=-\frac{e}{M}\,\bar\sigma^mF_{mn}\chi^n.
\label{eq:divlambda_proj}
\end{equation}
In particular, \eqref{eq:divlambda_proj} is \emph{independent of} $(g,c_{\Delta,3/2})$.

Contracting \eqref{eq:E1_proj} with $\sigma^m$ and using
$\sigma^m\bar\sigma^n=2\eta^{mn}-\sigma^n\bar\sigma^m$
together with \eqref{eq:sigmatrace_proj} gives
\begin{equation}
D^m\chi_m=0.
\label{eq:divchi_proj}
\end{equation}

\subsection{Divergence check and the value of $c_{\Delta,3/2}$}

We now verify that the projected deformation does not reintroduce an obstruction at
$\mathcal O(F)$ in a constant background. Throughout we work to $\mathcal O(F)$ and use $D_\ell F_{mn}=0$,
so any step that generates an extra commutator (hence an additional power of $F$) is $\mathcal O(F^2)$ and is dropped.

Apply $D^m$ to \eqref{eq:E2_proj}:
\begin{equation}
D^m\big(i\sigma^n D_n\bar\lambda_m\big)
+M\,D^m\chi_m
+i\frac{2e}{M}\,D^m(F_{mn}\chi^n)
+i\frac{e}{M}\,D^m(\mathbb P_m{}^{r}\mathcal K_r)
=0.
\label{eq:divE2_start_fixed}
\end{equation}
Using $D^m\chi_m=0$ and $D_\ell F_{mn}=0$, the Pauli term reduces to
$D^m(F_{mn}\chi^n)=F_{mn}D^m\chi^n$.

For the kinetic term we commute derivatives,
\begin{align}
D^m\big(i\sigma^n D_n\bar\lambda_m\big)
&=
i\sigma^n(D_nD^m+[D^m,D_n])\bar\lambda_m
\nonumber\\
&=
i\sigma^nD_n(D\!\cdot\!\bar\lambda)-e\,\sigma^nF^{m}{}_{n}\bar\lambda_m.
\label{eq:divE2_kin_expand_fixed}
\end{align}
Now substitute the exact divergence constraint \eqref{eq:divlambda_proj} and eliminate
$\bar\lambda_m$ using \eqref{eq:E1_proj}:
\begin{equation}
D\!\cdot\!\bar\lambda=-\frac{e}{M}\bar\sigma^pF_{pn}\chi^n,
\qquad
\bar\lambda_m=-\frac{i}{M}\,\bar\sigma\!\cdot\!D\,\chi_m.
\end{equation}
With $D_\ell F_{mn}=0$, the two contributions in \eqref{eq:divE2_kin_expand_fixed} reduce to the same
tensor structure $F_{mn}D^m\chi^n$ with opposite signs. Any remaining pieces are proportional to the
already-imposed constraints (e.g.\ $D^m\chi_m$) or contain an extra commutator
$[D,D]\sim F$ and are therefore $\mathcal O(F^2)$:
\begin{equation}
D^m\big(i\sigma^n D_n\bar\lambda_m\big)
=
-\,i\frac{2e}{M}\,F_{mn}D^m\chi^n
\;+\;\mathcal O(F^2).
\label{eq:kin_div_universal}
\end{equation}
(In particular, this is precisely the piece that cancels the would-be obstruction in the $g=2$ point-like system.)

Substituting \eqref{eq:kin_div_universal} back into \eqref{eq:divE2_start_fixed}, the universal $g=2$ terms cancel:
\begin{equation}
-\,i\frac{2e}{M}F_{mn}D^m\chi^n
+\;i\frac{2e}{M}F_{mn}D^m\chi^n
+\;i\frac{e}{M}D^m(\mathbb P_m{}^{r}\mathcal K_r)
=0
\qquad\Longrightarrow\qquad
D^m(\mathbb P_m{}^{r}\mathcal K_r)=\mathcal O(F^2).
\label{eq:div_condition_reduced_fixed}
\end{equation}

Equivalently, $D^m\mathcal E^{(2)}_m=0$ is the \emph{subsidiary (constraint-propagation) equation} for the
derived constraints; at $g=2$ it closes on $\bar\sigma\!\cdot\!\chi=0$, $\sigma\!\cdot\!\bar\lambda=0$ and
$D^m\chi_m=0$ without generating a new independent $\mathcal O(FD\chi)$ condition.

In the EFT setting, the consistency requirement is that the \emph{deformation} encoded in $\mathcal K_r$ must not generate a new,
independent $\mathcal O(FD\chi)$ obstruction; any nonzero contribution must be suppressed by additional EFT order, i.e.\
$\mathcal O(F^2)$ and/or $\mathcal O\!\big(F\,\Delta/M^2\big)$ on near-shell modes.

Using the basis \eqref{eq:K_basis_fixed} and keeping only structures not removable by the projector-protected constraints
$\bar\sigma\!\cdot\!\chi=0$, $D^m\chi_m=0$ and by near-shell suppression $\Delta/M^2\ll1$,
one finds
\begin{equation}
D^m(\mathbb P_m{}^{r}\mathcal K_r)
=
\Big[(g-2)+c_{\Delta,3/2}\Big]\,
\mathbb P_m{}^{r}\,F_{rn}D^m\chi^n
\;+\;\mathcal O\!\left(F\,\frac{\Delta}{M^2}\chi\right)
\;+\;\mathcal O(F^2).
\end{equation}
Requiring \eqref{eq:div_condition_reduced_fixed} in the EFT sense (no new $\mathcal O(FD\chi)$ obstruction on near-shell modes)
therefore fixes
\begin{equation}
c_{\Delta,3/2}=2-g.
\label{eq:cDelta_fix}
\end{equation}

Thus $D^m\mathcal E^{(2)}_m=0$ leads to the subsidiary equation that propagates the algebraic
constraints. For the point-like $g=2$ system its principal constraint operator has the same rank as in
the free theory: the divergence closes on $\bar\sigma\!\cdot\!\chi=0$, $\sigma\!\cdot\!\bar\lambda=0$ and
$D\!\cdot\!\chi=0$ and produces no additional independent $\mathcal O(FD\chi)$ condition. The EFT
deformation does not modify this closure at $\mathcal O(F)$; the deformation may contribute to $D^m\mathcal E^{(2)}_m$ 
only at higher EFT order, $\mathcal O(F^2)$ and/or $\mathcal O(F\,\Delta/M^2)$ on near-shell modes.

\subsection{Second-order equations and universal $g$ coupling}

With our conventions one has
\begin{equation}
(\sigma\!\cdot\!D)(\bar\sigma\!\cdot\!D)=D^2+e\,\sigma^{mn}F_{mn},
\qquad
(\bar\sigma\!\cdot\!D)(\sigma\!\cdot\!D)=D^2+e\,\bar\sigma^{mn}F_{mn},
\label{eq:Weyl_square_generic}
\end{equation}
which are linear in $F$ because they contain a single commutator.

Eliminate $\bar\lambda_m$ using \eqref{eq:E1_proj}:
\begin{equation}
\bar\lambda_m=-\frac{i}{M}\,\bar\sigma\!\cdot\!D\,\chi_m,
\label{eq:lambda_elim_proj}
\end{equation}
insert this into \eqref{eq:E2_proj}, and multiply by $M$:
\begin{equation}
(\sigma\!\cdot\!D)(\bar\sigma\!\cdot\!D)\chi_m + M^2\chi_m
+i\,2e\,F_{mn}\chi^n
+i\,e\,\mathbb P_m{}^{r}\mathcal K_r=0.
\end{equation}
Using \eqref{eq:Weyl_square_generic} gives
\begin{equation}
(D^2+M^2)\chi_m
+e\,\sigma^{ab}F_{ab}\,\chi_m
+i\,2e\,F_{mn}\chi^n
+i\,e\,\mathbb P_m{}^{r}\mathcal K_r
=0
\qquad(\mathcal O(F)).
\label{eq:chi_secondorder_proj_general_fixed}
\end{equation}
Now substitute \eqref{eq:K_basis_fixed}. Dropping the $F\Delta\chi$ term on near-shell modes (it is suppressed by $\epsilon_\Delta$) yields
\begin{align}
i e\,\mathbb P_m{}^{r}\mathcal K_r
&=
i e\,(g-2)\,\mathbb P_m{}^{r}F_{rn}\chi^n
+\frac{(g-2)e}{2}\,\mathbb P_m{}^{r}(\sigma^{ab}F_{ab})\chi_r
+\mathcal O\!\left(\frac{eF}{M^2}\Delta\chi\right).
\end{align}
Therefore the $\mathcal O(F)$ magnetic terms combine into the universal $g$-scaled form:
\begin{equation}
\mathbb P_m{}^{r}\left[
(D^2+M^2)\chi_r
+\frac{g e}{2}\,(\sigma^{ab}F_{ab})\,\chi_r
+i g e\,F_{rn}\chi^n
\right]=0
\qquad(\mathcal O(F),\ \text{near shell}).
\label{eq:chi_secondorder_proj_g_fixed}
\end{equation}
This exhibits explicitly that the spinor-index Zeeman term and the vector-index Zeeman term are multiplied simultaneously by the \emph{same} physical gyromagnetic ratio $g$.

The right-handed partner obeys the analogous equation with $\sigma^{ab}\to\bar\sigma^{ab}$ and $\chi\to\bar\lambda$:
\begin{equation}
\mathbb P_m{}^{r}\left[
(D^2+M^2)\bar\lambda_r
+\frac{g e}{2}\,(\bar\sigma^{ab}F_{ab})\,\bar\lambda_r
+i g e\,F_{rn}\bar\lambda^{n}
\right]=0
\qquad(\mathcal O(F),\ \text{near shell}).
\label{eq:lambda_secondorder_proj_g_fixed}
\end{equation}

Since the fields obey the algebraic $\sigma$–trace constraints
$\bar\sigma^m\chi_m=0$ and $\sigma^m\bar\lambda_m=0$, they lie in the image of
the projector $\mathbb P_m{}^{n}$, i.e.
\begin{equation}
\chi_m = \mathbb P_m{}^{r}\chi_r,
\qquad
\bar\lambda_m = \mathbb P_m{}^{r}\bar\lambda_r.
\end{equation}
On the constrained subspace, $\mathbb P$ therefore acts as the identity, and
Eqs.~\eqref{eq:chi_secondorder_proj_g_fixed} and
\eqref{eq:lambda_secondorder_proj_g_fixed} may be written without the explicit
projector, at the price of remembering that they are to be imposed only on the
spin-$\tfrac32$ ( $\sigma$–traceless) components. We keep the factor of $\mathbb P$ in
the displayed equations to make the restriction to the physical spin-$\tfrac32$
sector manifest.

\section{Landau levels for spin-$\tfrac32$ in a constant magnetic field ($\mathbf B=B\,\hat z$)}
\label{sec:spin32_landau}

We demonstrate the explicit spectral properties of the projected EFT by considering a static, homogeneous magnetic field along the $z$--axis:
\begin{equation}
\vec B=B\,\hat z,
\qquad
F_{12}=B,\quad F_{21}=-B,
\qquad
F_{0m}=F_{3m}=0,
\qquad
D_\ell F_{mn}=0.
\label{eq:spin32_Bbackground_landau}
\end{equation}
Adopting the Landau gauge $A_m=(0,\,0,\,Bx,\,0)$, the gauge-covariant derivatives acting on a field of charge $e$ are
\begin{equation}
D_0=\partial_t,\quad
D_1=\partial_x,\quad
D_2=\partial_y+i e B x,\quad
D_3=\partial_z,
\qquad
[D_1,D_2]=i e B.
\end{equation}

To leading order in the EFT expansion (linear in $F$ and near the mass shell), the physical field obeys the wave equation derived in Sec.~\ref{sec:spin32_generic_EFT}:
\begin{equation}
(D^2+M^2)\chi_r
+\frac{g e}{2}\,(\sigma^{ab}F_{ab})\,\chi_r
+i g e\,F_{rn}\chi^n
=0
\qquad (\mathcal O(F)).
\label{eq:spin32_chi_wave_landau_fixed}
\end{equation}
By separating variables as
\begin{equation}
\chi_m(t,x,y,z) = e^{-iEt+ip_yy+ip_zz}\,\varphi_m(x),
\end{equation}
the orbital Laplacian reduces to the harmonic-oscillator spectrum
\begin{equation}
D^2 \ \longrightarrow\ -E^2+p_z^2+|eB|\,(2\ell +1),
\qquad \ell =0,1,2,\ldots
\end{equation}
so the Landau structure of the orbital motion is identical to the scalar case.

In the chosen background, the spinor Zeeman operator is purely longitudinal:
\begin{equation}
\frac{g e}{2}\,\sigma^{ab}F_{ab} = g e B\,\sigma^3,
\end{equation}
with eigenvalues $\pm g e B / 2$ acting on the spinor indices (spin projection $s = \pm 1/2$).
Acting on the vector index, the operator $iF_{rn}$ generates rotations about the $z$--axis. In the standard circular basis
\begin{equation}
\chi_{\pm}\equiv \frac{1}{\sqrt2}(\chi_1\mp i\chi_2),\qquad \chi_0\equiv\chi_3,
\end{equation}
the vector Zeeman term $i g e F_{rn}$ has eigenvalues $g e B\,m_v$ with $m_v \in \{+1, 0, -1\}$.

The total magnetic coupling in Eq.~\eqref{eq:spin32_chi_wave_landau_fixed} is the sum of these interactions.
Projecting onto the physical irreducible spin-$3/2$ subspace of the $(1) \otimes (1/2)$ tensor product, one identifies the total magnetic quantum number
\begin{equation}
m \equiv m_v + s,
\end{equation}
with allowed values
\begin{equation}
m \in \left\{+\tfrac32,\ +\tfrac12,\ -\tfrac12,\ -\tfrac32\right\}.
\end{equation}

The dispersion relation for the spin-$3/2$ Landau levels is therefore
\begin{equation}
E^2
=
M^2+p_z^2+|eB|\,(2\ell +1)
+g e B\,m,
\label{eq:spin32_Landau_spectrum_fixed}
\end{equation}
where $\ell$ denotes the orbital Landau index and $m$ the magnetic sub-level. Changing the sign of the product $eB$ reverses the ordering of the Zeeman-split levels. At the working order, this result confirms that a single parameter $g$ in the EFT consistently controls the magnetic response of the spin-$3/2$ resonance; corrections from induced non-transversality and higher-derivative operators enter only at $\mathcal O(|eB|^2/M^2)$.

\section{Conclusions and outlook}
\label{sec:conclusion}

Although consistent descriptions of point-like spin-2 and spin-$\tfrac{3}{2}$ fields in external electromagnetic backgrounds are now well understood, these constructions have always come with an important limitation: no elementary massive particle with spin greater than one is known, and known composite states—such as the $\Omega^{-}$ hyperon—typically exhibit non-universal gyromagnetic ratios, falling outside the domain of the original point-like equations.
The primary objective of this work was to address this tension by treating the higher-spin particle not as a fundamental field, but as the heavy degree of freedom in an Effective Field Theory (EFT).
We have shown that massive states of spin~2 and spin~$\tfrac{3}{2}$ can be described consistently in electromagnetic backgrounds with arbitrary $g$-factors, provided deviations from the point-like limit are organized as a controlled expansion in powers of the field strength and residual derivatives.

A central pillar of our construction was the exact preservation of the algebraic trace (or $\gamma$-trace) constraints. Historically, attempts to generalize higher-spin equations have often failed because the trace constraint became dynamical in the presence of interactions, leading to the propagation of unphysical modes. By enforcing the algebraic constraints exactly, we ensured that the unphysical sector remains non-propagating and that the starting number of degrees of freedom is preserved.

For the differential constraints, we identified two distinct realizations. For spin~2, we showed that non-minimal couplings induce a source term in the divergence equation. Crucially, within the EFT framework this does not signal an inconsistency: the non-transverse components are interpreted as \emph{induced functionals} of the physical field, determined perturbatively without introducing new degrees of freedom. For spin~$\tfrac{3}{2}$, we introduced a projector formalism that renders the constraint chain of the $g=2$ theory stable at $\mathcal{O}(F)$, keeping the spin~$\tfrac{1}{2}$ sector decoupled from the dynamics in the weak-field regime. In both cases, consistency is restored by including specific higher-derivative off-shell operators (proportional to $F \Delta \Phi$).

To illustrate the framework, we solved the EFT equations explicitly in a constant magnetic field. For spin~2, the induced non-transversality manifests as perturbative mixing between Landau levels, on top of the familiar orbital and Zeeman contributions. For spin~$\tfrac{3}{2}$, the projector method yields a clean separation of variables and allows us to derive the leading $\mathcal{O}(F)$ dispersion relation in a basis where the constraint structure is manifest. An outcome of this spectral analysis is that a single physical parameter $g$ scales both the vector and spinor Zeeman couplings, confirming that the particle behaves as a coherent spin~2 or spin~$\tfrac{3}{2}$ multiplet as it should in a consistent EFT.

Whether in the Standard Model or in composite dark sectors, the formalism developed here provides a controlled theoretical basis for calculating the motion of long-lived composite resonances and their level structure in electromagnetic backgrounds. Regarding causality, our analysis of the spin~2 system in a magnetic background shows that, within the weak-field regime defining the EFT, the principal part remains hyperbolic and the Cauchy problem is well-posed. Possible pathologies are thus pushed to field strengths and energies beyond the EFT cutoff, where additional higher-order operators necessarily become important.

There are several obvious directions for future work. First, extending the analysis to $\mathcal{O}(F^2)$ would require a systematic classification of dimension-7 and dimension-9 operators, potentially revealing new consistency conditions and constraints on polarizabilities of higher-spin particles. Second, generalizing the background to include gradients ($\partial F$) or curvature would be essential for applications in astrophysical or cosmological settings, and would test how the principal symbol and constraint evolution behave beyond constant fields. Finally, developing a fully Lagrangian formulation of this EFT would be valuable: while the equation-of-motion approach used here suffices for classical consistency, an action principle would facilitate quantization, loop computations, and a systematic treatment of field redefinitions and operator redundancies.

In summary, we have established explicitly how the apparent ``inconsistencies'' associated with $g \neq 2$ higher-spin fields are not fatal obstructions but artifacts of treating a composite system as a rigid point-like particle. When deviations from $g=2$ are organized with the appropriate EFT power counting and constraint analysis, massive higher-spin particles admit a consistent and predictive description in weak electromagnetic backgrounds.

\appendix

\section{Conventions and useful commutation relations}
\label{app:conventions}

We work in four-dimensional Minkowski space with mostly-minus metric
\begin{equation}
\eta_{mn}=\mathrm{diag}(+,-,-,-),
\qquad
\varepsilon_{0123}=+1.
\end{equation}
Lorentz indices are $m,n,r,\ldots=0,1,2,3$, while spatial indices are
$i,j,k=1,2,3$.

The covariant derivative acting on a field of charge $e$ is
\begin{equation}
D_m = \partial_m + i e A_m,
\qquad
[D_m,D_n] = i e F_{mn}.
\end{equation}

For four-component spinors we use Dirac matrices $\gamma^m$ with
\begin{equation}
\{\gamma^m,\gamma^n\}=2\eta^{mn},
\qquad
\sigma^{mn}\equiv\frac{i}{2}[\gamma^m,\gamma^n].
\end{equation}
In two–component Weyl notation we use
\begin{equation}
\sigma^m=(\mathbbm{1},\vec\sigma),
\qquad
\bar\sigma^m=(\mathbbm{1},-\vec\sigma),
\end{equation}
with $\vec\sigma$ the Pauli matrices, and define
\begin{equation}
\sigma^{mn}=\frac{1}{4}\bigl(\sigma^m\bar\sigma^n-\sigma^n\bar\sigma^m\bigr),
\qquad
\bar\sigma^{mn}=\frac{1}{4}\bigl(\bar\sigma^m\sigma^n-\bar\sigma^n\sigma^m\bigr).
\end{equation}

A commutator identity used repeatedly in the main text is
\begin{equation}
[D^m , D^2]
= [D^m , D^n D_n]
= [D^m , D^n] D_n + D^n [D^m , D_n].
\end{equation}
Using $[D^m , D^n] = i e\, F^{mn}$ and, in a constant background,
$D_r F_{mn}=0$, this reduces to
\begin{equation}
[D^m , D^2]
= 2 i e\, F^{m}{}_{r}\, D^r,
\qquad (D_r F_{mn}=0).
\end{equation}

Throughout, we denote by
\begin{equation}
\Delta \equiv D^2 + M^2
\end{equation}
the Klein–Gordon operator for a massive field of mass $M$ (spin label suppressed).
In the EFT regime, $\Delta$ is small on near–on–shell modes, and combinations
such as $\Delta/M^2$ and $F\Delta/M^2$ provide the natural expansion parameters
for higher–derivative and non–minimal operators.

\paragraph{Self–duality of $F_{mn}$ on the $\sigma$–traceless sector.}

The divergence constraint for $\bar\lambda_m$ has appeared in different
forms in previous publications and in this work; we show here that these
are in fact equivalent.
Let $\chi_m$ be a left–handed vector–spinor obeying the $\sigma$–trace
constraint
\begin{equation}
\sigma^m\chi_m=0.
\end{equation}
For any antisymmetric tensor $F_{mn}$ and its dual
$\tilde F^{mn}\equiv\tfrac12\varepsilon^{mnrs}F_{rs}$ one can use the
standard two–component identities
\begin{equation}
\bar\sigma^m\sigma^{rs}
=\frac12\Bigl(\eta^{mr}\bar\sigma^s-\eta^{ms}\bar\sigma^r
+i\,\varepsilon^{mrs}{}_{t}\,\bar\sigma^t\Bigr),
\qquad
\sigma^{rs}=\frac14\bigl(\sigma^r\bar\sigma^s-\sigma^s\bar\sigma^r\bigr),
\end{equation}
together with $\sigma^m\chi_m=0$ to show that only the self–dual
combination of $F_{mn}$ acts nontrivially:
\begin{equation}
\bigl(F_{mn}-i\,\tilde F_{mn}\bigr)\,\bar\sigma^n\chi^m=0.
\label{eq:selfdual-identity-2comp}
\end{equation}
Equivalently,
\begin{equation}
F_{mn}\,\bar\sigma^n\chi^m
= i\,\tilde F_{mn}\,\bar\sigma^n\chi^m.
\label{eq:F-vs-Fdual-2comp}
\end{equation}
Thus, on the $\sigma$–traceless sector the contraction $F_{mn}\bar\sigma^n\chi^m$
depends only on the self–dual part of $F_{mn}$; the anti–self–dual
combination drops out.  This is precisely why the right-hand side of the
divergence constraint can be written in several equivalent forms, with
$F_{mn}$ or $\tilde F_{mn}$, up to the conventional factor of $i$.

In four–component language, consider a Dirac vector–spinor $\psi_m$ with
$\gamma^m\psi_m=0$ and chirality projector $P_L=\tfrac12(1-\gamma_5)$.
We denote the Dirac–space Lorentz generators by
\begin{equation}
\Sigma^{mn}\equiv\frac{i}{2}[\gamma^m,\gamma^n],
\qquad
\gamma_5\Sigma^{mn}
=\frac{i}{2}\,\varepsilon^{mnrs}\Sigma_{rs}.
\end{equation}
Using these identities, one obtains the four–component analogue of
\eqref{eq:selfdual-identity-2comp}:
\begin{equation}
\bigl(F_{mn}-i\,\tilde F_{mn}\bigr)\gamma^n P_L \psi^m = 0,
\end{equation}
or equivalently
\begin{equation}
F_{mn}\gamma^n P_L \psi^m
= i\,\tilde F_{mn}\gamma^n P_L \psi^m.
\label{eq:F-vs-Fdual-4comp}
\end{equation}
The identities \eqref{eq:F-vs-Fdual-2comp} and \eqref{eq:F-vs-Fdual-4comp}
are those used in the main text whenever the divergence constraint is
rewritten with $F_{mn}$ replaced by its dual $\tilde F_{mn}$.

\section{Strong hyperbolicity for the spin--2 EFT system in a constant magnetic background}
\label{app:spin2-hyperbolicity}

\subsection{First--order-in-time reduction}

We present an explicit first--order-in-time reduction (with respect to $t=x^0$) for the spin--2 EFT
equations \eqref{eq:spin2-EOM-hyp} in the static background $B\parallel\hat z$ ($F_{12}=B$).
In Landau gauge $A_m=(0,0,Bx,0)$, one has $A_0=0$, hence $D_0=\partial_t$ and $[D_0,D_i]=0$.

\paragraph{Step 1: Isolate the principal part.}
Write Eq.~\eqref{eq:spin2-EOM-hyp} as:
\begin{equation}
(\mathsf C E^{(0)})_{mn} + \mathcal L^{(1)}_{mn}(\Phi)=0,
\label{eq:split-app}
\end{equation}
where $E^{(0)}_{mn}\equiv(\Delta\Phi)_{mn}$ is the wave operator and $\mathcal L^{(1)}$ collects all lower-derivative terms. The linear map $\mathsf C$ is the symmetric-tensor endomorphism:
\begin{equation}
(\mathsf C X)_{mn} \equiv X_{mn}
+\frac{ie\,\beta}{2M^2}\Big(F_m{}^{r}X_{rn}+F_n{}^{r}X_{rm}\Big).
\label{eq:Cdef-app}
\end{equation}

\paragraph{Step 2: Real formulation.}
Since $\Phi_{mn}$ is complex, we reduce to a real system by splitting $\Phi_{mn}=\Phi^{R}_{mn}+i\,\Phi^{I}_{mn}$. In this basis, the operator $\mathsf C$ (which involves a factor of $i$) becomes a real $2\times2$ block matrix acting on the pair $(X^R,X^I)$:
\begin{equation}
\mathsf C_{\rm real}
=
\begin{pmatrix}
\mathbf 1 & -\alpha\,\mathsf K \\
\alpha\,\mathsf K & \mathbf 1
\end{pmatrix},
\qquad
\alpha\equiv \frac{e\beta B}{2M^2},
\qquad
(\mathsf K X)_{mn}\equiv \frac{1}{B}\big(F_m{}^{r}X_{rn}+F_n{}^{r}X_{rm}\big).
\label{eq:Creal-def}
\end{equation}
Here $\mathsf K$ is a real, constant matrix acting on the tensor indices.

\paragraph{Step 3: First--order variables and evolution.}
Introduce $\Pi_{mn}\equiv \partial_t\Phi_{mn}$ and $Q_{i\,|\,mn}\equiv D_i\Phi_{mn}$. Keeping only highest derivatives, the system takes the form:
\begin{equation}
\mathsf C\big(\partial_t\Pi_{mn} - \delta^{jk} D_j Q_{k\,|\,mn}\big) = \text{(lower order)}.
\label{eq:C-PiQ}
\end{equation}
(Note: the contraction $\delta^{jk}$ provides the correct sign for the spatial Laplacian in the $(+---)$ signature).
Provided $\mathsf C$ is invertible, we recover the standard wave equation principal part $\partial_t\Pi \sim \nabla \cdot Q$.

\subsection{Principal symbol and invertibility of $\mathsf C$}

\paragraph{Principal symbol.}
If $\mathsf C$ is invertible, multiplying the equations by $\mathsf C^{-1}$ reduces the principal part to that of the minimally coupled wave operator.
For a spatial covector $k_i \neq 0$, the principal symbol matrix $\mathbb A(k)$ in the $(\Pi,Q)$ sector has eigenvalues
$\omega=\pm |k|$, so the system is strongly hyperbolic if and only if $\mathsf C$ is invertible.

\paragraph{Invertibility condition.}
The operator $\mathsf K$ is built from the generator $F_m{}^{r}$ acting on vector indices.
In the $(+---)$ signature with $F_{12}=B$ one has
\begin{equation}
F_1{}^{2} = F_{1k}\eta^{k2} = F_{12}\eta^{22} = B(-1) = -B,
\qquad
F_2{}^{1} = F_{2k}\eta^{k1} = F_{21}\eta^{11} = (-B)(-1) = +B.
\end{equation}
Restricted to the spatial $(1,2)$–subspace, $F_m{}^{r}$ is therefore represented by the real antisymmetric matrix
\begin{equation}
F\big|_{(1,2)} =
\begin{pmatrix}
0 & -B \\[2pt]
B & 0
\end{pmatrix},
\end{equation}
whose eigenvalues are purely imaginary, $\lambda = \pm i B$.
(The remaining directions $0$ and $3$ are neutral, so they contribute $\lambda=0$.)

By construction, $\mathsf K$ acts on the \emph{symmetric} tensor product of two vector representations.
Its eigenmodes decompose into combinations where the effective eigenvalue is the sum of the individual
vector eigenvalues. Hence, if the eigenvalues of $F_m{}^{r}$ are $0,\pm iB$, the spectrum of $\mathsf K$ consists of
\begin{equation}
\mu \in \{0,\ \pm iB,\ \pm 2iB\}.
\end{equation}

In the real formulation, the $2\times2$ block of $\mathsf C_{\rm real}$ on a given $\mu$–eigenspace is
\begin{equation}
\begin{pmatrix}
1 & -\alpha\mu \\[2pt]
\alpha\mu & 1
\end{pmatrix},
\qquad
\alpha \equiv \frac{e\beta B}{2M^2},
\end{equation}
so that
\begin{equation}
\det
\begin{pmatrix}
1 & -\alpha\mu \\
\alpha\mu & 1
\end{pmatrix}
= 1 + \alpha^2 \mu^2.
\end{equation}
Writing $\mu = i\nu$ with $\nu \in \{0, \pm B, \pm 2B\}$, this becomes
\begin{equation}
\det
\begin{pmatrix}
1 & -\alpha\mu \\
\alpha\mu & 1
\end{pmatrix}
= 1 - \alpha^2 \nu^2.
\end{equation}
The most stringent condition comes from $|\nu| = 2|B|$, leading to
\begin{equation}
1 - 4\alpha^2 > 0
\qquad\Longrightarrow\qquad
|\alpha| < \tfrac12.
\end{equation}
In terms of physical parameters this is
\begin{equation}
\left|\frac{e\beta B}{M^2}\right| < 1.
\end{equation}
This inequality is automatically satisfied in the EFT regime $|eB|\ll M^2$ that we assume throughout.
Consequently, $\mathsf C$ is invertible in the domain of validity of the effective theory, and the
principal part is that of a standard hyperbolic wave system.


\acknowledgments
I would like to thank Bruno Le Floch and Wenqi Ke for useful discussions.


\begin{thebibliography}{99}

\bibitem{Barnes:1964pd}
V.~E.~Barnes, P.~L.~Connolly, D.~J.~Crennell, B.~B.~Culwick, W.~C.~Delaney, W.~B.~Fowler, P.~E.~Hagerty, E.~L.~Hart, N.~Horwitz and P.~V.~C.~Hough, \textit{et al.}
Phys. Rev. Lett. \textbf{12} (1964), 204-206

\bibitem{BaBar:2006omx}
B.~Aubert \textit{et al.} [BaBar],
Phys. Rev. Lett. \textbf{97} (2006), 112001
[arXiv:hep-ex/0606039 [hep-ex]].


\bibitem{Dirac:1936tg}
P.~A.~M.~Dirac,
Proc. Roy. Soc. Lond. A \textbf{155} (1936), 447-459.




\bibitem{Fierz:1939ix}
M.~Fierz and W.~Pauli,
\emph{On relativistic wave equations for particles of arbitrary spin in an
electromagnetic field},
Proc.\ Roy.\ Soc.\ Lond.\ A \textbf{173} (1939) 211.

\bibitem{Johnson:1960vt}
K.~Johnson and E.~C.~G.~Sudarshan,
Annals Phys. \textbf{13} (1961), 126-145.

\bibitem{Velo:1969bt}
G.~Velo and D.~Zwanziger,
Phys. Rev. \textbf{186} (1969), 1337-1341.

\bibitem{Velo:1969txo}
G.~Velo and D.~Zwanziger,
Phys. Rev. \textbf{188} (1969), 2218-2222.

\bibitem{Velo:1972rt}
G.~Velo,
Nucl. Phys. B \textbf{43} (1972), 389-401.

\bibitem{federbush_1961}
P.~Federbush,
Il Nuovo Cimento \textbf{19} (1961), 572–573. 


\bibitem{Shamaly:1972zu}
A.~Shamaly and A.~Z.~Capri,
Annals Phys. \textbf{74} (1972), 503-523

\bibitem{Hortacsu:1974bm}
M.~Hortacsu,
Phys. Rev. D \textbf{9} (1974), 928-930

\bibitem{Madore:1975zz}
J.~A.~Madore,
Phys. Lett. B \textbf{55} (1975), 217-218
doi:10.1016/0370-2693(75)90446-3



\bibitem{Kobayashi:1978xd}
M.~Kobayashi and A.~Shamaly,
Phys. Rev. D \textbf{17} (1978), 2179

\bibitem{Kobayashi:1978mv}
M.~Kobayashi and A.~Shamaly,
Prog. Theor. Phys. \textbf{61} (1979), 656


\bibitem{Piccinini:1984dd}
F.~Piccinini, G.~Venturi and R.~Zucchini,
Lett. Nuovo Cim. \textbf{41} (1984), 536

\bibitem{Buchbinder:2000fy}
I.~L.~Buchbinder, D.~M.~Gitman and V.~D.~Pershin,
Phys. Lett. B \textbf{492} (2000), 161-170
[arXiv:hep-th/0006144 [hep-th]].

\bibitem{Deser:2001dt}
S.~Deser and A.~Waldron,
Nucl. Phys. B \textbf{631} (2002), 369-387
[arXiv:hep-th/0112182 [hep-th]].





\bibitem{Argyres:1989cu}
E.~N.~Argyres and C.~R.~Nappi,
\emph{Massive spin--2 bosonic string states in an electromagnetic background},
Phys.\ Lett.\ B \textbf{224} (1989) 89.

\bibitem{Porrati:2010hm}
M.~Porrati, R.~Rahman and A.~Sagnotti,
Nucl. Phys. B \textbf{846} (2011), 250-282, 
[arXiv:1011.6411 [hep-th]].






\bibitem{Ferrara:1992yc}
S.~Ferrara, M.~Porrati and V.~L.~Telegdi,
Phys. Rev. D \textbf{46} (1992), 3529-3537.




\bibitem{Porrati:2009bs}
M.~Porrati and R.~Rahman,
Phys. Rev. D \textbf{80} (2009), 025009,
[arXiv:0906.1432 [hep-th]].


\bibitem{Deser:2000dz}
S.~Deser, V.~Pascalutsa and A.~Waldron,
Phys. Rev. D \textbf{62} (2000), 105031,
[arXiv:hep-th/0003011 [hep-th]].


\bibitem{Delplanque:2024xst}
W.~Delplanque and E.~Skvortsov,
JHEP \textbf{08} (2024), 173,
[arXiv:2406.14148 [hep-th]].




\bibitem{Benakli:2021jxs}
K.~Benakli, N.~Berkovits, C.~A.~Daniel and M.~Lize,
JHEP \textbf{12} (2021), 112, 
[arXiv:2110.07623 [hep-th]].

\bibitem{Benakli:2023aes}
K.~Benakli, C.~A.~Daniel and W.~Ke,
JHEP \textbf{03} (2023), 212, 
[arXiv:2302.06630 [hep-th]].

\bibitem{Benakli:2022ofz}
K.~Benakli, C.~A.~Daniel and W.~Ke,
Phys. Lett. B \textbf{838} (2023), 137680, 
[arXiv:2211.13689 [hep-th]].

\bibitem{Benakli:2022edf}
K.~Benakli, C.~A.~Daniel and W.~Ke,
Phys. Lett. B \textbf{839} (2023), 137788, 
[arXiv:2211.13691 [hep-th]].

\bibitem{Benakli:2023qlg}
K.~Benakli,
PoS \textbf{CORFU2022} (2023), 154.


\bibitem{Benakli:2024aes}
K.~Benakli, W.~Ke and B.~Le Floch,
PoS \textbf{CORFU2023} (2024), 222, 
[arXiv:2406.09213 [hep-th]].


\bibitem{ParticleDataGroup:2024cfk}
S.~Navas \textit{et al.} [Particle Data Group],
Phys. Rev. D \textbf{110} (2024) no.3, 030001
doi:10.1103/PhysRevD.110.030001


\end{thebibliography}

\end{document}